%% file: acl_latex.tex
\documentclass{article} 

\usepackage[final]{colm2026_conference}

\usepackage{microtype}
\usepackage{hyperref}
\usepackage{url}
\usepackage{booktabs}
\usepackage{wrapfig} 

\usepackage{xspace}
\usepackage{amsfonts}
\usepackage{times}
\usepackage[most]{tcolorbox} 
\usepackage{latexsym}
\usepackage{pifont}

\usepackage{amsmath} 
\usepackage{multirow}
\usepackage{makecell}
\usepackage[T1]{fontenc}
\usepackage[normalem]{ulem}

\usepackage{lineno}
\definecolor{darkblue}{rgb}{0, 0, 0.5}
\hypersetup{colorlinks=true, citecolor=darkblue, linkcolor=darkblue, urlcolor=darkblue}

\usepackage[utf8]{inputenc}
\usepackage{booktabs}
\usepackage{multirow}
\usepackage{colortbl}
\usepackage{tabularx}
\definecolor{mygray}{gray}{0.9}
\usepackage{makecell}
\usepackage{pifont} 
\usepackage{xspace}

\usepackage[ruled,vlined,linesnumbered]{algorithm2e}
\SetKwInput{KwIn}{Input}
\SetKwInput{KwOut}{Output}
\SetKwFunction{PairKeys}{PairKeys}
\SetKwFunction{ComponentKeys}{ComponentKeys}
\SetKwFunction{TopM}{Top-$M$}
\SetKwFunction{TopN}{Top-$N$}
\SetKwFunction{Neighbors}{Neighbors}
\SetKwFunction{Sim}{sim}

\usepackage{amsthm}

\newtheorem{definition}{Definition}

\usepackage{tikz}

\usepackage{microtype}

\usepackage{inconsolata}

\usepackage{graphicx}

\newcommand{\ours}{{\texttt{KAMR}}\xspace}
\title{KAMR: Grounding Generation via Knowledge-Aligned Multi-hop Retrieval}

\author{
Xiaochen Wang\textsuperscript{1},
Yuan Zhong\textsuperscript{1},
Haoyu Wang\textsuperscript{2},
Ting Wang\textsuperscript{3},
Fenglong Ma\textsuperscript{1}
\\[2pt]
\textsuperscript{1}The Pennsylvania State University
\\
\textsuperscript{2}University at Albany, State University of New York
\\
\textsuperscript{3}Stony Brook University, State University of New York
\\[2pt]
\texttt{\{xcwang,yfz5556,fenglong\}@psu.edu}
\\
\texttt{hwang28@albany.edu, twang@cs.stonybrook.edu}
}

\begin{document}
\maketitle
\begin{abstract}
Graph-based retrieval-augmented generation increasingly relies on multi-hop retrieval, where answering a query requires composing multiple connected knowledge-graph triplets.
However, existing retrievers often rank triplets independently via global semantic matching. Moreover, many multi-hop benchmarks provide only final answers, which limits supervision for query--triplet alignment and causes structurally necessary but weakly aligned facts to be missed. To address these issues, we propose a knowledge-aligned multi-hop retriever, {\ours}, which distinguishes anchor triplets that are strongly constrained by the query from connected triplets that are weakly aligned yet structurally linked to the anchors.
To mitigate the lack of query--triplet alignment supervision, we build a partial alignment dataset by masking triplet elements and prompting an LLM to generate corresponding queries, and optimize two contrastive objectives for pair-level and element-level matching.
At inference time, \ours retrieves anchors globally and then expands locally to collect connected evidence.
Across four benchmarks, three LLM backbones, and fourteen baselines, \ours consistently improves multi-hop retrieval and downstream question answering performance. 

\end{abstract}

\input{latex/intro_final}
\input{latex/related_work_v2}
\input{latex/method_final}
\input{latex/experiment}
\input{latex/case_study}

\input{latex/conclusion}

\clearpage

\section*{Acknowledgements}
This research was partially supported by a 2025/2026 Rising Researcher Grant from Penn State's Institute for Computational \& Data Sciences (RRID:SCR\_025154) and the National Science Foundation under Grant No. 2333790 and 2238275.


\bibliography{custom.bib}
\bibliographystyle{colm2026_conference}
\appendix

\input{latex/appendix}

\end{document}

%% file: latex/intro_final.tex
\section{Introduction}

Retrieval-Augmented Generation (RAG) systems increasingly rely on \textbf{multi-hop retrieval} to support complex reasoning over external knowledge sources. Unlike single-hop retrieval, which assumes that relevant evidence is directly aligned with the input query, multi-hop retrieval must cover multiple pieces of information that are only indirectly connected to the query and must be composed to produce correct answers. This setting naturally arises in knowledge graph (KG) reasoning, where answering a query often requires traveling multiple connected triplets. As a result, graph-based multi-hop retrieval has become a critical component for tasks such as multi-hop question answering and knowledge-grounded reasoning.

Previous work typically formulates graph-based multi-hop retrieval as conventional search. Many approaches retrieve triplets independently by globally matching the query against candidate evidence, using either sparse retrievers~\citep{hu2025cg, li2023graph} or dense retrievers~\citep{dong2023bridging, li2023graph, hu2025cg}. While effective in single-hop settings, such strategies struggle in multi-hop scenarios, where some required evidence has weak or no direct textual alignment with the query. Recent Graph RAG methods~\citep{hu2024grag, he2024g, li2024simple} partially address this issue by leveraging structural connectivity in the knowledge graph. Still, they largely rely on off-the-shelf semantic retrievers pretrained on generic text corpora, whose objectives focus on text-to-text matching rather than query--triplet alignment. Consequently, existing methods often fail to balance semantic relevance with structural compositionality, leading to either incomplete reasoning chains or the retrieval of noisy, weakly related evidence.

\begin{wrapfigure}{r}{0.62\linewidth}
    \vspace{-10pt}
    \centering
    \includegraphics[width=1\linewidth]{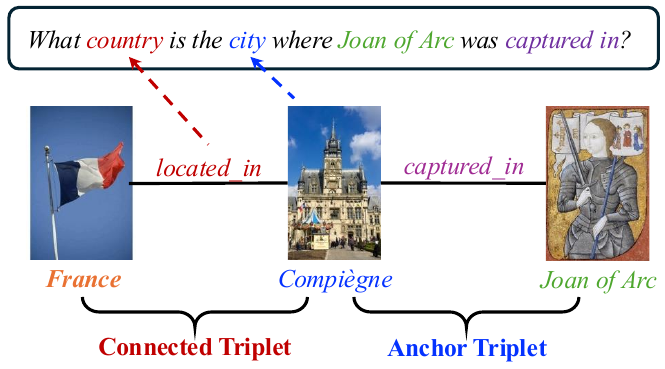}
    \vspace{-25pt}
    \caption{An example demonstrating the necessity of differentiating heterogeneous triplet types and explicitly modeling structural connectivity for multi-hop reasoning.}
    \label{fig:intro}
\end{wrapfigure}

At the core of this difficulty lies a fundamental challenge of graph-based multi-hop retrieval: queries are often underspecified with respect to the required triplets. In many cases, a query directly constrains only a part of the relevant evidence, while the remaining triplets are connected through intermediate entities or relations. 
As shown in Figure~\ref{fig:intro}, the query ``\textit{What country is the city where Joan of Arc was captured in?}'' can be answered by composing two triplets:
(\textit{Joan of Arc, captured in, Compi\`egne}) and (\textit{Compi\`egne, located in, France}).
The first triplet is strongly constrained by the entity-relation pair ``\textit{Joan of Arc}'' and ``\textit{captured in}'' and thus exhibits high semantic alignment with the query.
In contrast, the second triplet is connected to the query only through the intermediate entity ``\textit{{Compi\`egne}}'' and an implicitly relevant relation (``\textit{located in}''). Therefore, it exhibits weak textual alignment with the query and is unlikely to be retrieved by global semantic matching alone.
As a result, treating all triplets as equally aligned with the query is suboptimal. Instead, different triplets play different roles in the reasoning process and should be retrieved using different criteria. However, existing retrievers typically lack mechanisms to explicitly model such differences.

Another major challenge is pretraining graph-based multi-hop retrievers \textit{without query--triplet alignment supervision}. Although some methods have attempted to tune retrievers~\citep{li2024simple}, they typically assume access to explicit query--triplet alignment signals for supervision. While some benchmarks, such as PathQuestion~\citep{zhou2018interpretable}, provide alignment signals (e.g., gold reasoning paths), others, such as ComplexQuestions~\citep{bao2016constraint}, provide only final answers and do not explicitly annotate which knowledge triplets should be retrieved to support multi-hop reasoning. Thus, for such datasets, it is challenging to train retrievers that can distinguish between strongly aligned evidence and weakly aligned but structurally necessary evidence, which limits the generalizability of existing supervised methods~\citep{he2021improving}. Moreover, naive supervision at the triplet level ignores the fact that queries often specify only partial elements of a triplet, further exacerbating the mismatch between training objectives and inference requirements.

To address these challenges, we propose a novel \textbf{K}nowledge-\textbf{A}ligned \textbf{M}ulti-hop \textbf{R}etriever (\ours) that explicitly models partial query-triplet alignment and structural connectivity. Our key insight is to distinguish between two types of retrieval targets: \textbf{anchor triplets}, which are strongly constrained by the query and exhibit high semantic alignment, and \textbf{connected triplets}, which are weakly aligned with the query but are structurally connected to anchor triplets and essential for multi-hop reasoning. 

To mitigate the lack of query--triplet alignment supervision, we construct a partial alignment dataset by masking individual triplet elements and prompting a large language model (LLM) to generate corresponding queries. This allows us to pretrain {\ours} with two complementary contrastive objectives: a pair-level loss that aligns queries with partial triplets for anchor retrieval, and an element-level loss that aligns queries with individual triplet elements to support connected triplet retrieval. Importantly, this design avoids the need for explicit multi-hop supervision while aligning naturally with our two-stage inference procedure: we first retrieve anchor triplets, then expand to structurally connected triplets within their local neighborhoods to complete multi-hop evidence via graph connectivity.

Our contribution can be summarized as follows:
(1) We identify and formalize the distinction between anchor triplets and connected triplets in graph-based multi-hop retrieval, highlighting their different roles and alignment properties.
(2) We propose a novel graph-based multi-hop retrieval pretraining framework {\ours} that combines pair-level and element-level matching to balance semantic relevance and structural connectivity. 
(3) We construct a partial query-triplet alignment dataset that enables effective retriever pretraining without explicit multi-hop annotation.
(4) We demonstrate that the proposed \ours consistently improves multi-hop retrieval performance across four Graph RAG benchmarks on three LLM backbones, outperforming fourteen baselines.

%% file: latex/related_work_v2.tex
\section{Related Work}
\label{sec:related_work}

\noindent\textbf{Multi-hop Retrieval for Generation.}
Multi-step generation often requires evidence spanning multiple hops. Text-based RAG addresses this through iterative retrieval or query decomposition~\citep{jiang2023active,trivedi2023ircot,lee2024planrag,verma2024plan}, while Graph RAG retrieves structured evidence from knowledge graphs for more compositional reasoning~\citep{gao2023retrieval,guo-etal-2023-prompt,ma-etal-2023-query,min2019knowledge,zhang2025survey}. Recent methods also apply pretrained LM retrievers over graph elements in the conventional RAG setting~\citep{he2024g,li2023graph,li2024simple,hu2024grag}. In parallel, other works rely on task-specific generation-based reasoning, graph neural architectures, or iterative LLM prompting to induce reasoning paths or subgraphs~\citep{sun2019pullnet,atif2023beamqa,huang2021breadth,zhang2022subgraph,luo2023reasoning,sun2023think}. By contrast, we focus on retriever design in the conventional RAG setting, where the retriever is a lightweight component for efficient multi-hop evidence retrieval rather than a task-specific reasoning system~\citep{karpukhin2020dense,lewis2020retrieval}.

\noindent\textbf{Dense Retrievers for RAG.}
Dense retrievers link natural-language queries to external knowledge by ranking candidates via embedding-based similarity, forming the backbone of many RAG systems~\citep{robertson2009probabilistic,salton1988term,karpukhin2020dense,gao2021simcse,ni2022sentence}. Advanced variants can enhance semantic matching beyond exact token overlap and improve generation quality in RAG pipelines~\citep{chen2024bge,youdao_bcembedding_2023,gunther2023jina,shakir2024boost,formal2021splade,santhanam2022colbertv2,lin2023train, wang2026mkg}. While effective for plain-text RAG, these methods are tuned for unstructured text and overlook graph topology and relational dependencies needed for multi-hop reasoning.

\paragraph{KG-based Multi-hop Question Answering.}
KG-based multi-hop question answering has been extensively studied using task-specific models for direct answer prediction over diverse domain knowledge~\citep{wang2025medmkg,craciun2025graf}. Representative approaches include question-specific subgraph construction \citep{sun2018open}, iterative retrieval and reasoning \citep{sun2019pullnet, wang2026gpr}, graph neural reasoning \citep{sorokin2018modeling,shi2021transfernet,mavromatis2022rearev}, and path-based reasoning \citep{saxena2020improving}. However, most of these methods formulate retrieval and answering as part of a unified, task-specific QA model trained end-to-end with explicit answer supervision. This setting is fundamentally different from the multi-hop retrieval problem in Graph RAG that we study, where the goal is to improve the retrieval of supporting facts by optimizing an independent retriever in an unsupervised manner, without tuning the backbone LLM. Therefore, although these studies are relevant, we do not include them in direct comparison because they fall outside the scope of our problem.

%% file: latex/method_final.tex
\section{The Proposed \ours Framework}

\subsection{Task Definition}
In contrast to existing graph-based RAG methods that retrieve triplets independently, multi-hop retrieval emphasizes \textbf{structural connectivity} among the retrieved triplets. In this work, we use connectivity as an \emph{operational} multi-hop constraint: it is necessary for compositional reasoning but not sufficient on its own. Thus, we define graph-based multi-hop retrieval as follows:
\begin{definition}[Graph-based Multi-hop Retrieval]\label{def:task}
    Let $\mathcal{G} = (\mathcal{E}, \mathcal{R}, \mathcal{T}_{\mathcal{G}})$ be a knowledge graph, where $\mathcal{E}$ is the entity set, $\mathcal{R}$ is the relation set, and $\mathcal{T}_{\mathcal{G}} \subseteq \mathcal{E} \times \mathcal{R} \times \mathcal{E}$ is the set of KG triplets.
    Given a textual query $q$, the objective of graph-based multi-hop retrieval is to select a subset $\mathcal{T}_q \subset  \mathcal{T}_{\mathcal{G}}$ that is semantically relevant to $q$, i.e.,
    \[
        \mathcal{T}_q = \text{Retriever}(q, \mathcal{G}),
    \]
    and satisfies the following multi-hop connectivity constraint:
    \begin{equation*}
    \forall (h_i, r_i, t_i) \in \mathcal{T}_q,\; \exists (h_j, r_j, t_j) \in \mathcal{T}_q \setminus \{(h_i, r_i, t_i)\}\;\;\text{s.t.}\;\; \{h_i, t_i\} \cap \{h_j, t_j\} \neq \emptyset,\quad |\mathcal{T}_q| = K \ll |\mathcal{T}_{\mathcal{G}}|.
    \end{equation*}
    where $h_i, t_i \in \mathcal{E}$ and $r_i \in \mathcal{R}$.
\end{definition}

\subsection{Targets of Multi-hop Retrieval}
\label{sec:target}
An ideal graph-based multi-hop retriever should return a set of triplets $\mathcal{T}_q$ that are not only well aligned with the query $q$, but also provide the missing information required to answer it. In the multi-hop setting, however, these two objectives often conflict, as queries are frequently underspecified and do not fully constrain all relevant facts.
This observation motivates targeted mechanisms that explicitly account for query–triplet misalignment in multi-hop retrieval. Accordingly, we introduce two categories of retrieval targets that we use throughout the paper based on the degree to which a triplet is constrained by the current context.


\begin{definition}[Anchor Triplets]\label{def:anchor_triplets}
Let $\mathcal{S}_q$ denote the set of entities and relations extracted from the query $q$.
A triplet $\tau_i = (h_i, r_i, t_i) \in \mathcal{T}_{\mathcal{G}}$ is called an
\textbf{anchor triplet} if it matches exactly two elements extracted from the query $q$, i.e.,
\[
\tau_i \in \mathcal{A}_q
\quad \text{s.t.} \quad
\left| \mathcal{S}_q \cap \text{Elem}(\tau_i) \right| = 2,
\]
where $\mathcal{A}_q$ denotes the set of anchor triplets for $q$, and $\text{Elem}(\tau_i)$ represents the element set of $\tau_i$.
\end{definition}

\begin{definition}[Connected Triplets]\label{def:connected_triplets}
A triplet $\tau_j$ is called a \textbf{connected triplet} if it matches exactly one element extracted from the query and is \emph{graph-connected} to the retrieved evidence. Formally, let $\mathcal{C}_q$ be the \emph{smallest} set such that for any $\tau_j=(h_j,r_j,t_j)\in\mathcal{T}_{\mathcal{G}}$,
\begin{gather*}
\tau_j \in \mathcal{C}_q, \quad \text{s.t.} \quad \left| \mathcal{S}_q \cap \text{Elem}(\tau_j) \right| = 1,\;\;\text{and} \\
\exists\, \tau_p\in \mathcal{A}_q \cup \mathcal{C}_q \;\; \text{s.t.} \;\;
\{h_p, t_p\} \cap \{h_j, t_j\} \neq \emptyset.
\end{gather*}
This definition allows a connected triplet to attach to an anchor triplet or to another connected triplet, enabling multi-hop expansion beyond two hops.
\end{definition}

Figure~\ref{fig:intro} shows an anchor triplet \textit{Joan of Arc, captured in, Compi\`egne}. It also shows a connected triplet \textit{Compi\`egne, located in, France}. This connected triplet contains an implicitly relevant relation (``\textit{located in}'') and is graph-connected to the retrieved evidence through the shared entity ``\textit{Compi\`egne}''.
Anchor triplets are typically easier to retrieve due to their strong semantic alignment with the query, but they are insufficient for multi-hop reasoning on their own. Connected triplets complement anchors by progressively introducing new, structurally linked evidence, supporting multi-hop chains of arbitrary length within the retrieval budget.

\subsection{\ours Pretraining}
Training a multi-hop retriever poses significant challenges for two main reasons. First, there is a lack of query–triplet alignment supervision. For instance, although benchmarks such as Complex Web Questions (CWQ)~\citep{talmor2018web} have been widely adopted for knowledge graph-based question answering, they provide only final answers and do not include explicit annotations indicating which triplets should be retrieved during reasoning.

Second, as discussed above, multi-hop reasoning requires retrieving both anchor and connected triplets, yet the query typically specifies only partial elements of the relevant triplets. Consequently, effective retrieval cannot rely on matching the query against fully specified triplets. Instead, the retriever must model the correspondence between the textual query and \textbf{partially specified triplets}, which substantially complicates pretraining and supervision.

To address these challenges, we first use LLMs to construct a training dataset for multi-hop retriever pretraining, and then introduce a tailored pretraining loss that accounts for partial query–triplet alignment.

\subsubsection{Partial Alignment Dataset Construction}
Given a complete triplet $\tau_i =(h_i, r_i, t_i) \in \mathcal{T}_{\mathcal{G}}$, we generate three partial triplets by masking one element at a time, yielding the set, 
\begin{equation}\label{eq:partial_triplet}
    \mathcal{U}(\tau_i) = \{(h_i, r_i), (r_i, t_i), (h_i, t_i)\}.
\end{equation}
Each partial triplet can be converted into a natural language query by prompting an LLM\footnote{The prompt is provided in Appendix~\ref{app:prompt}.}.

Let $u_i = (e_i^1, e_i^2) \in \mathcal{U}(\tau_i)$ denote a partial triplet, where $e_i^k \in \{h_i, r_i, t_i\}$ and $k=1,2$. $q_i$ denotes the corresponding query generated from $u_i$. We construct a pretraining dataset $\mathcal{U} = \{(q_i, u_i)\}_{i=1}^{|\mathcal{U}|}$, where $|\mathcal{U}|$ is the total number of partial query–triplet alignment pairs. This dataset, detailed in Appendix~\ref{app:implementation}, is used to pretrain the multi-hop retriever.

\subsubsection{Pretraining Loss Design}
Since graph-based multi-hop retrieval aims to retrieve both anchor and connected triplets, we define the overall pretraining objective as the sum of two loss terms:
\begin{equation}\label{eq:loss}
    \mathcal{L} = \mathcal{L}_{a} + \mathcal{L}_{c},
\end{equation}
where $\mathcal{L}_{a}$ corresponds to the pretraining loss for anchor triplets, and $\mathcal{L}_{c}$ corresponds to the pretraining loss for connected triplets. We next describe the formulation of each loss term in detail.

\noindent \textbf{Anchor Triplet Retrieval Pretraining.}
Given a partial query-triplet alignment pair $(q_i, u_i)$, we first encode the query and the corresponding partial triplet using a query encoder $\text{ENC}_q(\cdot)$ and a triplet-element encoder $\text{ENC}_u(\cdot)$ as follows:
\begin{equation}\label{eq:encoder}
    \mathbf{q}_i = \text{ENC}_{q}(q_i), \;\; \mathbf{u}_i = \text{ENC}_{u}(e_i^1 \oplus e_i^2),
\end{equation}
where $\oplus$ denotes concatenation. We then optimize their alignment using the InfoNCE loss~\citep{oord2018representation}:
\begin{equation}\small
\label{eq:anchor_loss}
\mathcal{L}_{a}
= - \frac{1}{|\mathcal{B}|} \sum_{i=1}^{|\mathcal{B}|}
\log
\frac{\exp(\mathrm{sim}(\mathbf{q}_i, \mathbf{u}_i)/T)}
{\sum_{j=1}^{|\mathcal{B}|} \exp(\mathrm{sim}(\mathbf{q}_i, \mathbf{u}_{j})/T)},
\end{equation}
where $\mathcal{B}\subseteq\mathcal{U}$ denotes the mini-batch of training samples, $T$ is a temperature hyperparameter, and the remaining partial triplets within the batch serve as negatives.

\noindent \textbf{Connected Triplet Retrieval Pretraining.}
When constructing the partial alignment dataset, we do not explicitly create multi-hop supervision pairs for connected triplet pretraining. This is because the partial query--triplet alignment samples are sufficient to support the pretraining of both anchor and connected triplets. By Definition~\ref{def:connected_triplets}, a connected triplet overlaps with the query on only one element, while it must connect to previously retrieved evidence through a shared endpoint entity. Consequently, connected retrieval requires identifying the query-relevant element under structural constraints.

This observation motivates us to directly model \textit{element-level similarity} between the query $q_i$ and individual triplet elements, rather than relying on explicit multi-hop supervision during pretraining. Accordingly, we define another InfoNCE loss to pretrain the retriever for connected-triplet retrieval as follows:
\begin{equation}\small
\label{eq:expand_loss}
\mathcal{L}_{c}
= - \frac{1}{2|\mathcal{B}|} \sum_{i=1, k=1}^{|\mathcal{B}|,2}
\log
\frac{\exp(\mathrm{sim}(\mathbf{q}_i, \mathbf{e}_i^{k})/T)}
{\sum_{j=1,k=1}^{|\mathcal{B}|,2} \exp(\mathrm{sim}(\mathbf{q}_i, \mathbf{e}_{j}^{k})/T)},
\end{equation}
where $\mathbf{e}_i^{k} = \text{ENC}_u(e_i^k)$ is the embedding of the $k$-th element $e_i^k \in u_i$ $(k= 1, 2)$.

\subsection{{\ours} Inference}
\label{sec:inference}
After optimizing the pretraining objective in Eq.~\eqref{eq:loss} on the constructed dataset $\mathcal{U}$, we directly apply the pretrained graph-based multi-hop retriever to a variety of downstream tasks. During inference, the retriever first identifies \emph{anchor triplets} via the pair-level matching learned from anchor-triplet pretraining. It then performs \emph{iterative multi-hop expansion}: at each step, it retrieves new \emph{connected triplets} that are graph-connected to the \emph{current retrieved set}, including both anchors and previously retrieved connected triplets. This iterative strategy generalizes the original two-hop setting to arbitrary hops under a fixed retrieval budget.

\noindent \textbf{Anchor Triplet Retrieval.} 
Each triplet $\tau_i \in \mathcal{T}_{\mathcal{G}}$ is first converted into a set of partial triplets $\mathcal{U}(\tau_i)$ according to Eq.~\eqref{eq:partial_triplet}. We then compute the cosine similarity between the query embedding $\mathbf{q}$ and the embedding of each partial triplet in $\mathcal{U}(\tau_i)$, where embeddings are obtained using Eq.~\eqref{eq:encoder}. The maximum similarity score among the three partial query–triplet pairs is taken as the overall similarity between the query $q$ and the triplet $\tau_i$. Finally, we retrieve the top-$M$ triplets as anchor candidates, i.e., $|\mathcal{A}_q|=M$. 

\noindent \textbf{Connected Triplet Retrieval.}
We perform connected-triplet retrieval via iterative expansion from the current retrieved set. Let $\mathcal{T}_q^{(0)}=\mathcal{A}_q$ be the initial anchors. We run up to $N$ expansion iterations; at iteration $\ell\ge 1$, we extract the $g$-hop neighborhood of the current evidence $\mathcal{T}_q^{(\ell-1)}$ from the knowledge graph $\mathcal{G}$ and construct the candidate set $X\big(\mathcal{T}_q^{(\ell-1)}\big)$ according to Definition~\ref{def:connected_triplets}. Each candidate is graph-connected to previously retrieved evidence and overlaps with the query on exactly one element.

For each candidate triplet $x\in X\big(\mathcal{T}_q^{(\ell-1)}\big)$, we identify a connecting triplet in $\mathcal{T}_q^{(\ell-1)}$ that shares an endpoint entity with $x$ and remove this shared entity from $x$. For example, if $x=(h,r,t)$ is connected to the current evidence via $h$, we score $x$ using the remaining two elements $r$ and $t$. We embed the remaining two elements using the element encoder $\text{ENC}_u(\cdot)$, and score $x$ by the maximum cosine similarity between the query embedding $\mathbf{q}$ and these element embeddings. This inference-time scoring is aligned with the element-level pretraining objective in Eq.~\eqref{eq:expand_loss}. We then add the top-ranked candidates to form $\mathcal{T}_q^{(\ell)}$, ensuring $|\mathcal{T}_q^{(\ell)}|\le K$.
We stop when the retrieval budget $K$ is reached, $N$ iterations are completed, or no new candidates can be added. The final connected-triplet set is $\mathcal{C}_q = \mathcal{T}_q \setminus \mathcal{A}_q$.

\noindent \textbf{Final Multi-hop Evidence.}
We return the retrieved multi-hop evidence set in the form required by Definition~\ref{def:task}:
\begin{equation}
\label{eq:union_new}
\mathcal{T}_{q} = \mathcal{A}_{q} \cup \mathcal{C}_{q},
\end{equation}
where $|\mathcal{A}_q| = M$ and the total retrieval budget is $|\mathcal{T}_{q}| = K$. Accordingly, $|\mathcal{C}_q| = K-M$.

This inference strategy is well aligned with our pretraining objectives: anchor retrieval relies on pair-level matching to address partial query–triplet alignment, while connected retrieval leverages element-level matching under anchor-induced structural constraints to recover weakly aligned yet composable multi-hop evidence.
Algorithm~\ref{alg:twostage} in Appendix~\ref{app:alg} summarizes the overall retrieval procedure. 

%% file: latex/experiment.tex
\section{Experiments}
\label{sec:experiment}

\subsection{Experiment Settings}

\noindent\textbf{Datasets.}
We evaluate \ours in a staged manner, from multi-hop retrieval quality to multi-hop retrieval-based applications.
For the retrieval quality evaluation, we use PathQuestion (\textbf{PQ-2H} for questions depending on 2-hop facts and \textbf{PQ-3H} for those relying on 3-hop triplets)~\citep{zhou2018interpretable}, which provides gold reasoning paths and enables explicit measurement of whether a retriever can recover the supporting triplets.
To evaluate downstream utility, we further consider question answering on two datasets with complementary characteristics: PathQuestion, which features a single answer and fixed 2- and 3-hop reasoning chains, and Complex Web Questions (\textbf{CWQ})~\citep{talmor2018web}, which allows multiple valid answers and requires up to 4-hops of reasoning.\footnote{Additional experiment results on LC-QUAD~\citep{trivedi2017lc} datasets are available in Appendix~\ref{app:exp}.}

\noindent\textbf{Backbones.}
Our retrieval strategies are tested with several LLM backbones pretrained on general-domain corpora, including ChatGPT-3.5 Turbo~\citep{achiam2023gpt}, LLaMA2-7B~\citep{touvron2023llama}, and Qwen3-8B~\citep{yang2025qwen3}. These models span different scales and include both open- and closed-source LLMs, ensuring a comprehensive evaluation across architectures.

\noindent\textbf{Baselines.}
We compare \ours against four categories of retrievers. 
(1) \textit{Lexical-based} methods include BM25~\citep{robertson2009probabilistic} and TF-IDF~\citep{sparck1972statistical}. 
(2) \textit{Semantic-based} dense retrievers include DistilBERT~\citep{Sanh2019DistilBERTAD}, BGE~\citep{chen2024bge}, BCE~\citep{youdao_bcembedding_2023}, Jina~\citep{gunther2023jina}, MXBAI~\citep{shakir2024boost}, SPLADE~\citep{formal2021splade}, ColBERTv2~\citep{santhanam2022colbertv2}, Dragon~\citep{lin2023train}, and Hybrid~\citep{li2023graph}. 
(3) \textit{Structure-based} graph retrievers include G-RAG~\citep{hu2024grag} and G-Retriever~\citep{he2024g}. 
(4) \textit{Graph-pretraining} method SKP~\citep{dong2023bridging}. 
Detailed configurations and implementation for all baselines as well as {\ours} are provided in Appendix~\ref{app:implementation}.



\begin{table*}[t]
\centering
\caption{Retrieval and generation performance on PathQuestion. The best and second-best results per backbone and dataset in the generation part are highlighted in \textbf{bold} and \underline{underlined}, respectively.}
\label{tab:pq_retrieval}
\resizebox{\textwidth}{!}{%
\begin{tabular}{l|l|cc|cc|cc|cc|cc}
\toprule
\multirow{3}{*}{\makecell[c]{\textbf{Retriever}\\\textbf{Category}}}
& \multirow{3}{*}{\textbf{Method}}
& \multicolumn{4}{c|}{\textbf{Retrieval}}
& \multicolumn{6}{c}{\textbf{Generation}} \\
\cline{3-12}
&
& \multicolumn{2}{c|}{\textbf{PQ-2H}}
& \multicolumn{2}{c|}{\textbf{PQ-3H}}
& \multicolumn{2}{c|}{\textbf{Qwen3-8B}}
& \multicolumn{2}{c|}{\textbf{LLaMA2-7B}}
& \multicolumn{2}{c}{\textbf{ChatGPT-3.5}} \\
\cline{3-12}
&
& \makecell[c]{\textbf{Triplet}\\\textbf{Recall}} & \makecell[c]{\textbf{Path}\\\textbf{Recall}}
& \makecell[c]{\textbf{Triplet}\\\textbf{Recall}} & \makecell[c]{\textbf{Path}\\\textbf{Recall}}
& \textbf{PQ-2H} & \textbf{PQ-3H}
& \textbf{PQ-2H} & \textbf{PQ-3H}
& \textbf{PQ-2H} & \textbf{PQ-3H} \\
\midrule

\makecell[c]{\textbf{\ding{55}}}
& {LLM-only}
& -- & -- & -- & --
& 46.09 & 33.54
& 12.47 & 8.50
& 39.47 & 26.24 \\
\midrule

\multirow{2}{*}{\makecell[c]{\textbf{Lexical}\\\textbf{Based}}}
& BM25
& 63.47 & 26.94 & 63.48 & 40.32
& 42.66 & 56.21
& 48.32 & 49.77
& 51.83 & 53.89 \\
& TF-IDF
& 63.78 & 27.57 & 64.45 & 40.52
& 42.08 & 55.19
& 46.70 & 49.25
& 50.10 & 52.23 \\
\midrule

\multirow{9}{*}{\makecell[c]{\textbf{Semantic}\\\textbf{Based}}}
& BCE
& 77.83 & 55.66 & 72.03 & 49.13
& 57.04 & 57.63
& 56.18 & 48.53
& 60.80 & 52.09 \\
& BGE
& 83.49 & 66.98 & 76.73 & 55.89
& 63.65 & 60.39
& 60.32 & 50.98
& 67.35 & 54.15 \\
& ColBERT
& 80.37 & 62.16 & 72.60 & 48.15
& 59.04 & 56.64
& 59.06 & 50.08
& 62.78 & 54.17 \\
& DistilBERT
& 34.83 & 17.77 & 27.98 & 10.04
& 24.60 & 30.17
& 44.70 & 31.72
& 45.70 & 37.18 \\
& Dragon
& 85.95 & 71.91 & 78.71 & 59.10
& 65.96 & 59.77
& 62.57 & 50.59
& 68.37 & 54.32 \\
& Hybrid
& 85.88 & 71.75 & 79.18 & 59.35
& 64.99 & 58.75
& 61.79 & 50.27
& 67.16 & 54.42 \\
& Jina
& 82.34 & 65.04 & 74.18 & 51.31
& 62.68 & 58.23
& 60.48 & 48.90
& 65.46 & 52.75 \\
& MXBAI
& 10.59 & 1.47 & 6.25 & 0.50
& 9.51 & 7.27
& 31.13 & 19.93
& 38.20 & 26.01 \\
& SPLADE
& 80.16 & 60.32 & 75.07 & 53.10
& 58.96 & 60.10
& 60.64 & 52.13
& 63.42 & 54.02 \\
\midrule

\multirow{2}{*}{\makecell[c]{\textbf{Structure}\\\textbf{Based}}}
& G-RAG
& 92.92 & 86.90 & 59.04 & 30.92
& \underline{72.40} & 51.21
& \underline{68.61} & 46.33
& \underline{74.16} & 50.00 \\
& G-Retriever
& 14.47 & 9.91 & 4.14 & 2.81
& 12.05 & 5.96
& 47.59 & 25.00
& 46.12 & 27.74 \\
\midrule

\multirow{2}{*}{\makecell[c]{\textbf{Graph}\\\textbf{Pretraining}}}
& SKP
& 75.68 & 57.76 & 53.26 & 31.05
& 50.79 & 43.52
& 60.06 & 41.22
& 66.35 & 45.15 \\
& \ours ($N=1$)
& \underline{96.36} & \underline{92.87} & \underline{85.60} & \underline{67.06}
& \textbf{73.53} & \underline{62.50}
& \underline{68.61} & \textbf{52.98}
& \underline{75.86} & \underline{56.54} \\
& \ours ($N=2$)
& \textbf{99.50} & \textbf{99.16} & \textbf{94.09} & \textbf{85.78}
& 69.70 & \textbf{68.69}
& \textbf{68.69} & \underline{52.53}
& \textbf{78.25} & \textbf{59.81} \\
\bottomrule
\end{tabular}
}
\vspace{-0.2in}
\end{table*}

\subsection{Multi-hop Retrieval Quality Evaluation}
\label{sec:retr_eval}
We explicitly evaluate retrieval quality on PathQuestion (PQ-2H and PQ-3H) using two metrics under a fixed retrieval budget of $K=50$ triplets. Under the same budget, we compare two settings that vary the number of expansion iterations $N$ in our inference procedure: $N=1$ and $N=2$.
\textbf{Triplet Recall} measures whether each gold triplet appearing in the annotated reasoning path is retrieved within the budget.
\textbf{Path Recall} measures whether the retriever recovers the entire reasoning path, i.e., all triplets required to form the gold multi-hop chain.\footnote{Note that we retrieve two different types of triplets, making it impossible to use other ranking metrics such as nDCG in the evaluation.}

Table~\ref{tab:pq_retrieval} reports retrieval performance on PathQuestion under the same retrieval budget. We observe that \ours achieves the best results on both PQ-2H and PQ-3H, with particularly strong Path Recall, indicating more stable recovery of complete multi-hop chains. Moreover, increasing the number of expansion iterations from $N=1$ to $N=2$ improves performance, especially on PQ-3H. This suggests that multi-round iterative expansion helps retrieve connected triplets that are farther away from the anchor triplets in the knowledge graph, improving multi-hop coverage.

Even the stronger baselines still fall short of a consistently high Path Recall, especially on PQ-3H, indicating that recovering complete multi-hop chains remains challenging under a fixed retrieval budget.
In addition, we observe two representative failure cases among baselines.
MXBAI is primarily a reranker and typically requires a lexical retriever (e.g., Solr~\citep{shahi2015apache}) for coarse filtering, so using it alone leads to many irrelevant candidates.
G-Retriever’s heuristic prize-based subgraph extraction optimizes the overall subgraph score, which can introduce non-essential triplets and dilute evidence-critical retrieval.

\begin{table*}[t]
\centering
\caption{Evaluation Results on CWQ Dataset.}
\label{tab:cwq}
\resizebox{\textwidth}{!}{%
\begin{tabular}{l|l|cccc|cccc|cccc}
\toprule
\multirow{2}{*}{\makecell[c]{\textbf{Retriever}\\\textbf{Category}}} & \multirow{2}{*}{\textbf{Method}} &
\multicolumn{4}{c|}{\textbf{Qwen3-8B}} &
\multicolumn{4}{c|}{\textbf{LLaMA2-7B}} &
\multicolumn{4}{c}{\textbf{ChatGPT-3.5}} \\
\cline{3-14}
& & \textbf{Acc} & \textbf{Prec} & \textbf{Rec} & \textbf{F1}
  & \textbf{Acc} & \textbf{Prec} & \textbf{Rec} & \textbf{F1}
  & \textbf{Acc} & \textbf{Prec} & \textbf{Rec} & \textbf{F1} \\
\midrule

\makecell[c]{\textbf{\ding{55}}}
& {LLM-only}
& 25.99 & 29.88 & 24.42 & 25.49
& 28.23 & 9.99  & 28.23 & 14.70
& 37.88 & 42.19 & 35.63 & 38.63 \\
\midrule

\multirow{2}{*}{\makecell[c]{\textbf{Lexical}\\\textbf{Based}}}
& BM25
& 20.44 & 24.02 & 19.55 & 20.44
& 29.41 & 34.81 & 28.09 & 29.38
& 31.70 & 36.22 & 30.05 & 31.27 \\
& TF-IDF
& 25.83 & 29.76 & 24.51 & 25.54
& 33.26 & 38.69 & 31.41 & 32.82
& 35.84 & 40.92 & 33.71 & 35.13 \\
\midrule

\multirow{9}{*}{\makecell[c]{\textbf{Semantic}\\\textbf{Based}}}
& BCE
& 32.36 & 36.48 & 29.57 & 30.96
& 36.11 & 41.74 & 33.72 & 35.31
& 40.28 & 44.77 & 37.00 & 38.57 \\
& BGE
& 35.59 & 39.45 & 32.72 & 34.08
& 38.35 & 44.21 & 35.85 & 37.55
& 42.90 & 47.24 & 39.35 & 40.95 \\
& ColBERT
& 36.48 & 40.27 & 33.05 & 34.49
& 37.31 & 42.96 & 34.82 & 36.44
& 42.66 & 47.52 & 39.33 & 40.99 \\
& DistilBERT
& 19.22 & 23.76 & 18.27 & 19.30
& 27.38 & 32.57 & 26.12 & 29.01
& 30.85 & 36.34 & 29.60 & 32.64 \\
& Dragon
& \underline{37.25} & \underline{41.22} & \underline{34.12} & \underline{35.53}
& 38.70 & \underline{44.80} & \underline{36.23} & \underline{37.93}
& \underline{44.45} & \underline{49.22} & \underline{40.91} & \underline{42.57} \\
& Hybrid
& 36.95 & 40.44 & 33.66 & 32.02
& 35.12 & 40.70 & 33.25 & 36.61
& 39.55 & 44.52 & 37.06 & 40.47 \\
& Jina
& 34.00 & 37.92 & 31.16 & 32.50
& 36.16 & 42.34 & 33.98 & 35.64
& 40.94 & 45.82 & 37.59 & 39.25 \\
& MXBAI
& 17.23 & 20.56 & 16.21 & 17.05
& 25.58 & 30.70 & 24.24 & 25.49
& 32.39 & 37.81 & 30.76 & 32.16 \\
& SPLADE
& 34.94 & 38.40 & 31.70 & 33.02
& 37.52 & 43.08 & 34.79 & 36.41
& 43.45 & 47.64 & 39.39 & 41.04 \\
\midrule

\multirow{2}{*}{\makecell[c]{\textbf{Structure}\\\textbf{Based}}}
& G-RAG
& 29.17 & 33.29 & 26.40 & 27.76
& 34.19 & 39.88 & 31.88 & 33.47
& 31.87 & 36.44 & 30.55 & 33.21 \\
& G-Retriever
& 27.87 & 31.42 & 25.65 & 26.80
& 38.37 & 43.56 & 35.73 & 37.30
& 33.51 & 39.42 & 31.84 & 34.99 \\
\midrule

\multirow{2}{*}{\makecell[c]{\textbf{Graph}\\\textbf{Pretraining}}}
& SKP
& 29.37 & 32.16 & 27.56 & 29.67
& 28.54 & 33.48 & 27.04 & 29.86
& 33.43 & 38.71 & 31.71 & 34.83 \\
& \ours ($N=1$)
& \textbf{38.71} & \textbf{42.40} & \textbf{34.90} & \textbf{36.43}
& \textbf{40.19} & \textbf{45.40} & \textbf{36.87} & \textbf{38.93}
& \textbf{45.10} & \textbf{49.59} & \textbf{41.09} & \textbf{42.70} \\
& \ours ($N=2$)
& 32.86 & 36.59 & 30.03 & 31.35
& \underline{39.17} & 44.77 & 36.22 & \underline{37.93}
& 44.35 & 49.11 & 40.56 & 42.27 \\
\bottomrule
\end{tabular}
}
\vspace{-20pt}
\end{table*}





\subsection{Generation Evaluation}
\label{sec:qa_eval}
Tables~\ref{tab:pq_retrieval} and~\ref{tab:cwq} summarize question answering evaluation results on PathQuestion and CWQ datasets across three LLM backbones.
We use accuracy as the evaluation metric for the single-answer dataset PathQuestion, and accuracy, precision, recall, and F1 scores for the multi-answer dataset CWQ.
Overall, \ours consistently achieves the best performance across all three backbones, indicating that the proposed alignment-oriented training and structure-aware inference yield stable gains across diverse generators and evaluation settings. On CWQ, notably, {\ours} with $N=2$ can underperform $N=1$ under the fixed budget, because allocating more budget to connected triplets reduces the number of anchor triplets and may hurt performance when anchor ranking is already challenging; this highlights the need to balance anchor and connected triplets in practice.


Aligning with the retrieval trends in Table~\ref{tab:pq_retrieval}, on PathQuestion, structure-aware retrieval can be effective when the required evidence is relatively local.
G-RAG performs competitively on PQ-2H, which aligns with its design of retrieving and encoding a 1-hop subgraph as evidence.
However, its performance degrades notably on PQ-3H, where the required reasoning chain extends beyond a single local neighborhood.
In contrast, \ours maintains more competitive performance on PQ-3H, suggesting that integrating anchors with connected triplets better retrieves longer compositional chains, notably facilitating the grounded generation. 

The QA task is more challenging on the CWQ dataset due to mixed reasoning depths, multiple correct answer candidates, and noisier query phrasing and entity mentions.
In this setting, strong semantic retrievers, e.g., Dragon, remain competitive because semantic matching is more tolerant to lexical variation and noisy mentions.
Nevertheless, \ours still achieves the best overall results across all backbones, suggesting that targeted pretraining better bridges query--triplet mismatch under noisy inputs, while the inference strategy enables broader evidence acquisition that can support multiple correct answers.


By combining partial-alignment-oriented pretraining with a structure-aware inference strategy, \ours better balances relevance and connectivity during retrieval, yielding stable, backbone-agnostic gains on both PathQuestion and CWQ datasets. Extra experiments for question answering are available at Appendix~\ref{app:exp}.

%% file: latex/case_study.tex
\subsection{Ablation Study}

\begin{wraptable}{r}{0.40\columnwidth}
\vspace{-30pt}
\centering
\caption{Ablation Study of {\ours}.}
\label{tab:ablation}
\resizebox{\linewidth}{!}{%
\begin{tabular}{l |l |c}
\toprule
 \textbf{Setting}& \textbf{Model} & \textbf{Accuracy} \\
\midrule
Full Model& ${\ours}$                          & \textbf{75.86} \\\midrule
\multirow{2}{*}{\makecell[l]{Pretraining\\Ablation}} 
& Without $\mathcal{L}_{\text{a}}$   & 48.79 \\
& Without $\mathcal{L}_{\text{c}}$   & 75.10 \\
\midrule
\multirow{3}{*}{\makecell[l]{Inference\\Ablation}} 
& Anchor Only        & 63.00 \\
& Full Anchor Triplets  & 75.26 \\
& Full Connected Triplets      & 74.79 \\
\bottomrule
\end{tabular}
}
\vspace{-15pt}
\end{wraptable}

To quantify the contribution of each design choice in \ours, we conduct ablations on the PQ-2H dataset with ChatGPT-3.5-Turbo and report accuracy in Table~\ref{tab:ablation}, with $N$ configured to 1. 
Each variant is constructed to directly correspond to a specific component in our methodology, allowing us to isolate its effect on end-to-end generation.

\textbf{Ablating the pretraining objectives.}
Our pretraining jointly handles partial alignment for anchor triplets $\mathcal{L}_a$ (i.e., Eq.~\eqref{eq:anchor_loss}) and connected triplets $\mathcal{L}_c$ (i.e., Eq.~\eqref{eq:expand_loss}), mirroring the two retrieval targets in Section~\ref{sec:target}.
To assess their roles, we remove each pretraining objective individually.
Excluding $\mathcal{L}_{a}$ leads to a severe degradation, consistent with our formulation that accurate anchor identification is the prerequisite for inducing a reliable local search region and subsequent expansion.
Removing $\mathcal{L}_{c}$ also hurts performance, indicating that learning element-level alignment is necessary for selecting informative connected triplets, given the limited partial overlap between semantics in the query and connected triplet.

\textbf{Ablating the connected triplet retrieval in inference.}
A key component of \ours is the connected triplet retrieval module, which complements globally retrieved anchors by adding structurally connected triplets from their local neighborhoods.
To test its necessity during inference, we remove the connected triplet retrieval module and keep only the anchor triplets, denoted ``\textit{Anchor Only}''.
This variant exhibits clear degradation, confirming that local expansion is essential for acquiring intermediate evidence that is weakly aligned with the query but required to complete multi-hop reasoning chains.

\textbf{Ablating partial alignment design in inference.}
As discussed in Section~\ref{sec:inference}, we use only partial triplets during inference, with $N$ configured to 1. We evaluate this design using two variants. First, for anchor retrieval, instead of computing similarity between the query and partial triplets, we use full triplets; this ablation is denoted ``\textit{Full Anchor Triplets}'' in Table~\ref{tab:ablation}, and we observe a drop in accuracy. Second, for connected-triplet retrieval, we use full triplets instead of element-level similarity; this variant is denoted ``\textit{Full Connected Triplets}'', and performance also drops.
Together, these ablations show that \ours has a redundancy-free design, offering a clearer view of the principles that drive its performance.



\subsection{Qualitative Study}

Beyond the quantitative results, we present two qualitative analyses to provide an intuitive understanding of how {\ours} retrieves evidence and how its structure-aware search supports downstream generation.

\begin{figure*}[t]
    \centering
    \begin{minipage}[t]{0.48\textwidth}
        \centering
        \includegraphics[width=\linewidth]{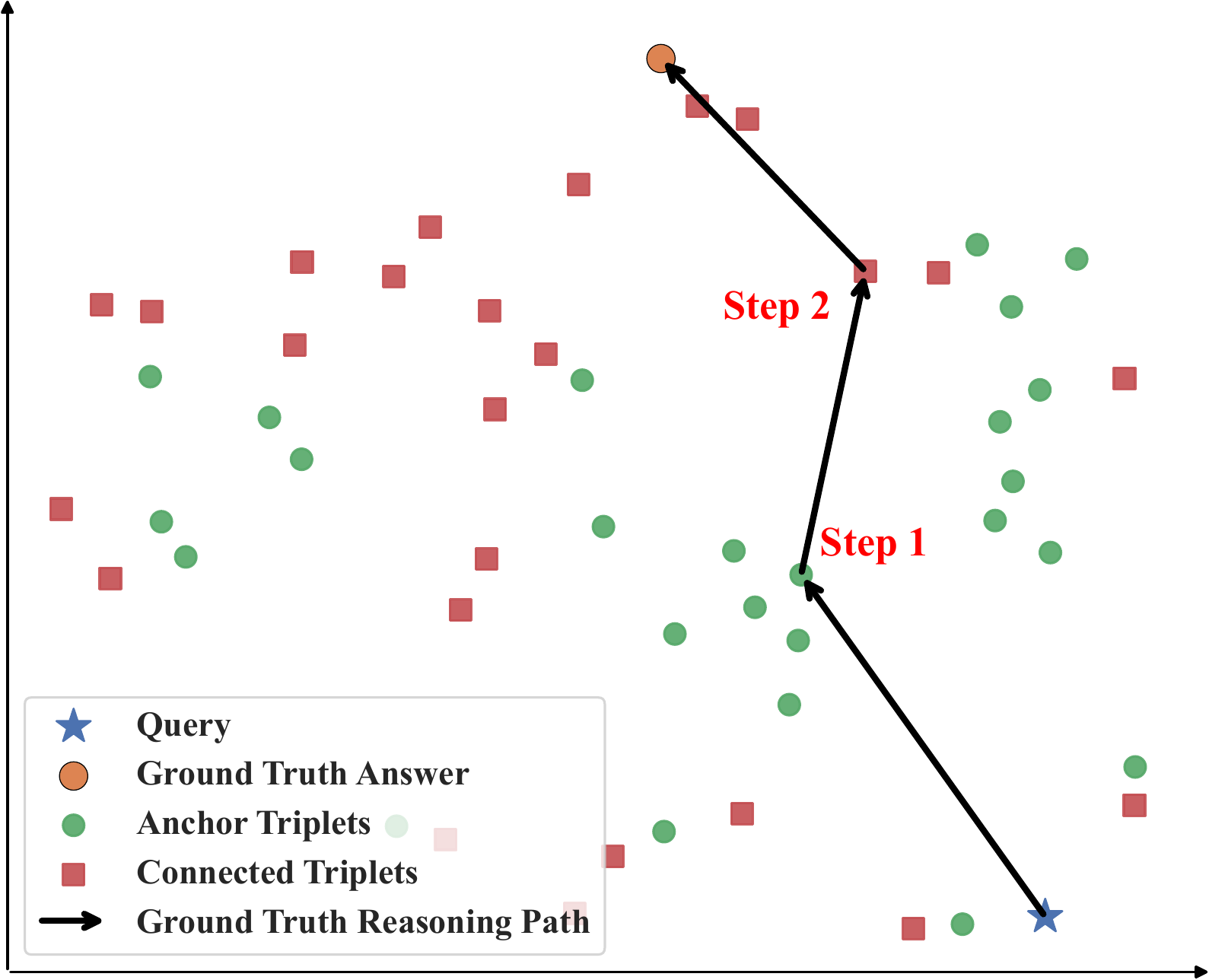}
        \vspace{-2pt}
        {\small (a) Visualization of how {\ours} retrieves semantically distant yet structurally relevant triplets.}
    \end{minipage}\hfill
    \begin{minipage}[t]{0.52\textwidth}
        \centering
        \includegraphics[width=\linewidth]{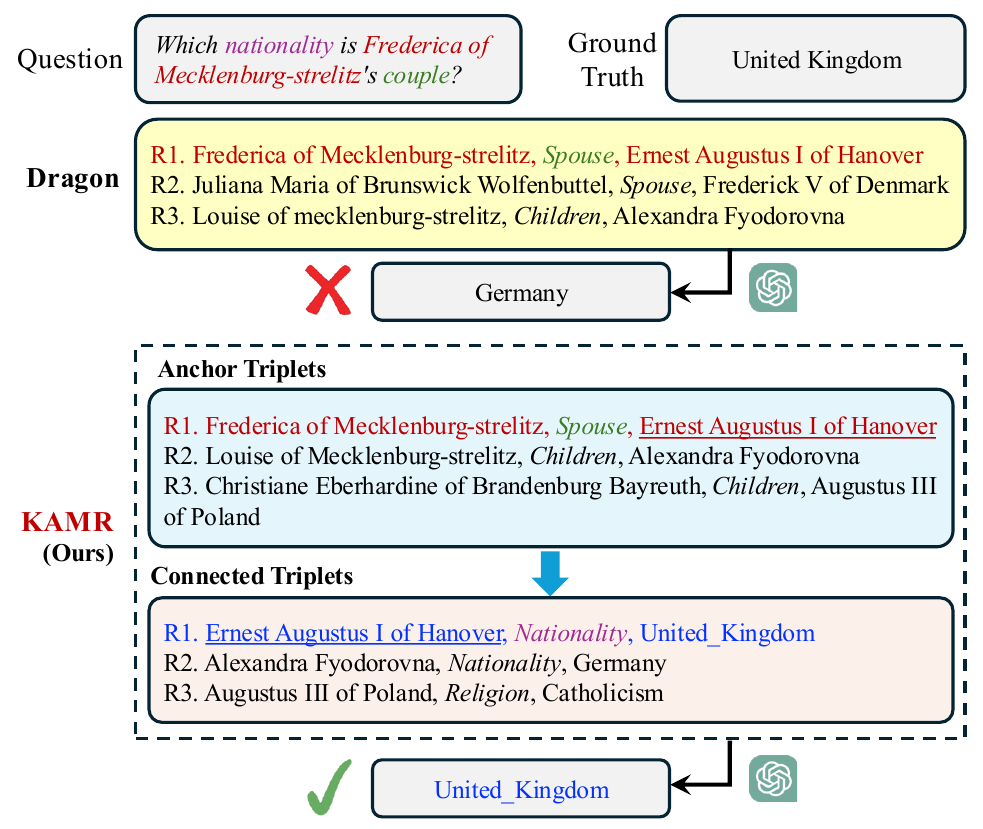}
        \vspace{-2pt}
        {\small (b) Case study comparison.}
    \end{minipage}
    \vspace{-10pt}
    \caption{Qualitative study.}
    \label{fig:vis}
    \vspace{-20pt}
\end{figure*}

\textbf{Case study of retrieved evidence.}
We examine a natural-language query from PQ-2H and compare the evidence retrieved by \ours with that of a strong baseline retriever, Dragon, as shown in Figure~\ref{fig:vis}(b). While a semantic retriever often captures the anchor triplet (\textit{Frederica of Mecklenburg-strelitz, Spouse, Ernest Augustus I of Hanover}), it may miss weakly aligned but structurally necessary supporting triplets such as (\textit{Ernest Augustus I of Hanover, Nationality, United\_Kingdom}), leading to fragmented evidence and less grounded answers. In contrast, \ours retrieves such facts by expanding from anchor triplets through graph connectivity, producing a more complete evidence set and more accurate generation.



\textbf{Semantic investigation via visualization.}
To further isolate the role of connected triplet retrieval, we conduct a semantic-space analysis using t-SNE~\citep{maaten2008visualizing} on another set of retrieval examples in PQ-2H.
Figure~\ref{fig:vis}(a) visualizes the query, the retrieved anchor triplets $\mathcal{A}_q$, and the expanded triplets $\mathcal{C}_q$.
As expected, when retrieval relies primarily on query semantics, the model tends to select anchor triplets that lie close to the query embedding.
As a complement, searching connected triplets with graph structure brings in additional triplets that may be farther away semantically, but are connected through the graph and contain knowledge-rich intermediate information.
These connected triplets acquired through {\ours} help bridge missing hops in the reasoning chain toward the final answer, thereby improving the stability and completeness of the evidence used for generation.




%% file: latex/conclusion.tex
\section{Conclusion}
\label{sec:conclusion}

We presented \ours, a retriever for multi-hop fact acquisition over knowledge graphs that addresses partial query--triplet alignment and structural compositionality. \ours constructs training data via LLM-guided triplet masking and query generation, and applies contrastive pretraining with pair-level and element-level objectives. At inference time, \ours retrieves anchor triplets globally and expands locally to collect connected evidence. Experiments show consistent gains in multi-hop retrieval and downstream question answering.

%% file: latex/appendix.tex




\section{LLM Usage Statement}

Large language models (LLMs) are a central object of study in this paper. Our research investigates how LLMs can be better supported by multi-hop retrieval for graph-based augmentation and generation, and we evaluate their behavior and outcomes within this retrieval-augmented framework. In this sense, LLMs are not merely auxiliary tools but an integral component of the problem setting and experimental analysis.

Separately, we used GPT-5 only for manuscript editing. Its use was limited to surface-level refinement, including improving phrasing, correcting grammar, and enhancing readability. All scientific contributions, including the problem formulation, methodology, experimental design, results, and interpretations, were developed by the authors.

All LLM-involved outputs were reviewed for accuracy and appropriateness. The authors take full responsibility for the validity and integrity of the final content.

\section{More Information on {\ours}}


\subsection{Prompt for Pretraining Dataset Construction}
\label{app:prompt}

The prompt for data construction from triplets is listed in Figure~\ref{fig:prompt}.

\subsection{Algorithm of {\ours}}
\label{app:alg}
Algorithm~\ref{alg:twostage} provides an intuitive overview of the full inference pipeline of {\ours}, facilitating a clear understanding and reliable reproducibility.

\begin{figure}[h]
\centering
\begin{tcolorbox}[
  title={Prompt for Pretraining Dataset Construction},
  colback=white,
  colframe=blue!50!black
]
\small
You are given a triplet from a knowledge graph in the form of (head, relation, tail), but one element is missing, shown as `??'.\par
\vspace{0.6\baselineskip}
Your job is to generate a single natural language question based on the incomplete triple, such that the missing part is what the question is asking for.\par
\vspace{0.6\baselineskip}
The question should be a complete sentence ending with a question mark.\par
\vspace{0.6\baselineskip}
Do NOT generate multiple questions. Stop after the first question. Do NOT include the masked answer or the triple in your output.\par
\vspace{0.6\baselineskip}

Triple: \{triplet\}\par
Masked Answer: \{answer\}\par
Output:
\end{tcolorbox}
\vspace{-10pt}
\caption{Prompt for Pretraining Dataset Construction.}
\label{fig:prompt}
\end{figure}

\section{Experiment Details}
\label{app:setting}


\subsection{Baselines} 

For a controlled and comparable evaluation under the conventional RAG setting, we focus on lightweight retrievers that can be deployed as an efficient standalone retrieval module. Methods that require end-to-end training with task-specific supervision or rely on heavy multi-step inference with large LLM prompting are not included, since their computation and optimization protocols are not directly comparable to fixed-budget retrieval and would confound the retrieval-focused evaluation. To this extent, we include the following graph-retrieval baselines for comparison:

\uline{\textit{BM25}}~\citep{robertson2009probabilistic} is a classic lexical retriever based on the probabilistic relevance framework. It ranks documents by aggregating exact term matches between the query and document, with a saturated term-frequency component and explicit document-length normalization (controlled by hyperparameters such as $k_1$ and $b$), making it a strong sparse baseline for keyword-style retrieval.

\uline{\textit{TF-IDF}}~\citep{sparck1972statistical} is a sparse vector-space retriever that represents each document (and query) with term weights computed from term frequency (TF) and inverse document frequency (IDF). It retrieves documents by comparing these weighted bag-of-words vectors (commonly via cosine similarity), favoring terms that are frequent in a document but rare across the corpus, and serving as a widely used baseline for lexical matching.

\uline{\textit{G-Retriever}}~\citep{he2024g} is a retrieval-augmented generation framework designed for question answering over textual graphs. It retrieves relevant nodes and edges based on semantic similarity and constructs subgraphs using the Prize-Collecting Steiner Tree (PCST)~\citep{goemans1995general} algorithm to form concise, query-relevant subgraphs for generation.

\uline{\textit{G-RAG}}~\citep{hu2024grag} is a graph retrieval-augmented generation method that enhances LLMs by retrieving and integrating textual subgraphs. It represents subgraphs as pooled embeddings of k-hop ego-graphs and retrieves them to incorporate both textual and topological information through dual prompting, improving performance on multi-hop reasoning tasks.

\uline{\textit{Hybrid}}~\citep{li2023graph} is a hybrid retrieval model that combines sparse retrieval (BM25) and dense retrieval (DPR) for coarse retrieval, followed by reranking with a cross-encoder to improve retrieval performance.

\uline{\textit{SKP}}~\citep{dong2023bridging} leverages traditional approaches like contrastive learning and masked language prediction on graphs to obtain a more graph-concentrated encoder for retrieval, enhancing the model's ability to represent complex subgraphs.

\uline{\textit{BGE}}~\citep{chen2024bge} is a versatile embedding-based retrieval model that supports multiple languages and tasks. It leverages dense, sparse, and multi-vector modalities to offer strong generalization across domains and text granularities.

\uline{\textit{BCE (BCEmbedding)}}~\citep{youdao_bcembedding_2023} is a bilingual/cross-lingual embedding framework optimized for efficient first-stage retrieval and (optional) reranking. It is designed to work well when dealing with mixed‐language content, providing embeddings that capture both semantic and cross‐language signals.

\begin{algorithm}[t]
\caption{Two-Stage Multi-hop Retrieval at Inference Time}\small
\label{alg:twostage}
\DontPrintSemicolon
\KwIn{Query $q$; knowledge graph $\mathcal{G}$; encoders $\text{ENC}_q(\cdot),\text{ENC}_u(\cdot)$; hop radius $g$; budgets $M,K$; expansion iterations $N$}
\KwOut{Multi-hop evidence set $\mathcal{T}_q$}

\BlankLine
\textbf{Encode query:} $\mathbf{q}\leftarrow \text{ENC}_q(q)$\;

\BlankLine
\textbf{Stage 1: Anchor triplet retrieval}\;
\ForEach{$\tau=(h,r,t)\in \mathcal{T}_{\mathcal{G}}$}{
    $\mathcal{U}(\tau)\leftarrow \{(h,r),(r,t),(h,t)\}$\;
    $s_A(q,\tau)\leftarrow \max\limits_{u=(e^1,e^2)\in \mathcal{U}(\tau)}\Sim\!\big(\mathbf{q},\text{ENC}_u(e^1\oplus e^2)\big)$\;
}
$\mathcal{A}_q \leftarrow \TopM\big(\{(\tau,s_A(q,\tau)):\tau\in\mathcal{T}_{\mathcal{G}}\}\big)$\;

\BlankLine
\textbf{Stage 2: Iterative connected triplet retrieval}\;
$\mathcal{T}_q \leftarrow \mathcal{A}_q$\;
\For{$\ell \leftarrow 1$ \KwTo $N$}{
    \If{$|\mathcal{T}_q| \ge K$}{\textbf{break}\;}
    $X(\mathcal{T}_q) \leftarrow \{x\in \Neighbors(\mathcal{T}_q,g): x \text{ satisfies Definition~\ref{def:connected_triplets}}\}$\;
    \ForEach{$x\in X(\mathcal{T}_q)$}{
        \textbf{Find} $\tau_p\in \mathcal{T}_q$ connected to $x$ via a shared endpoint entity $e_s$\;
        $\mathcal{E}_x \leftarrow \text{Elem}(x)\setminus \{e_s\}$\;
        $s_C(q,x)\leftarrow \max\limits_{e\in \mathcal{E}_x}\Sim\!\big(\mathbf{q},\text{ENC}_u(e)\big)$\;
    }
    $\Delta \leftarrow \TopN\big(\{(x,s_C(q,x)):x\in X(\mathcal{T}_q)\}\big)$\;
    $\mathcal{T}_q \leftarrow \mathcal{T}_q \cup \Delta$\;
}
$\mathcal{C}_q \leftarrow \mathcal{T}_q \setminus \mathcal{A}_q$\;

\BlankLine
\textbf{Compose evidence:}\;
$\mathcal{T}_q \leftarrow \mathcal{A}_q \cup \mathcal{C}_q$\;

\Return $\mathcal{T}_q$\;
\end{algorithm}

\uline{\textit{Jina Embeddings}}~\citep{gunther2023jina} are embedding models provided by Jina that aim for high retrieval performance through rich sentence/text representations. They focus on scalability, multi-lingual support, and ease of integration into existing retrieval pipelines.

\uline{\textit{MXBAI}}~\citep{shakir2024boost} is a retrieval/embedding family that emphasizes high performance on embedding benchmarks (e.g. sentence embeddings) and aims to balance between embedding quality and computational efficiency. It supports downstream tasks such as semantic search and retrieval.

\uline{\textit{SPLADE}}~\citep{formal2021splade} is a sparse representation model that transforms text into high-dimensional sparse vectors, preserving interpretability and allowing efficient matching with traditional retrieval indexing (e.g., inverted indices) while still capturing semantic similarity beyond keywords.

\uline{\textit{ColBERT}}~\citep{santhanam2022colbertv2} is a late-interaction retrieval model which encodes queries and documents/token sequences into token-level embeddings. It retains fine-grained interaction at search time, but optimizes memory/storage and inference through compression and efficient matching (e.g. MaxSim), giving strong accuracy across both in-domain and out-of-domain benchmarks.

\uline{\textit{DistilBERT}}~\citep{Sanh2019DistilBERTAD} is a distilled, lighter-weight version of BERT. It has fewer layers and parameters, enabling faster inference and lower resource use, and is often used as a baseline encoder for retrieval tasks where speed and efficiency matter.

\subsection{Implementation}
\label{app:implementation}

All experiments were conducted on four NVIDIA A6000 GPUs with CUDA version 12.0, running on Ubuntu 20.04.6 LTS. 
Pretraining is performed for 5 epochs using AdamW~\citep{loshchilov2017decoupled} with a batch size of 512 and a learning rate of $2\times 10^{-5}$.

For pretraining dataset construction, we derive triplets from Freebase~\citep{bollacker2008freebase}, restricting them to entities linked to the CWQ and PathQuestion datasets while keeping the synthetic corpus independent of retrieval evaluation and downstream question answering to prevent data leakage. We employ LLaMA-3.1-8B-Instruct~\citep{grattafiori2024llama} during this process, yielding 840,875 synthesized natural-language queries, which, together with their corresponding partial triplets, form the pretraining dataset. For methods that involve retriever pretraining (SKP and {\ours}), pretraining is conducted on the same synthetic dataset.

During inference, the 1-hop neighborhood of each anchor triplet is used as the search space for expansion.
\(\mathcal{A}_{q}\) and \(\mathcal{C}_{q}\) are each set to size \(25\) by choosing $M=25$ and $N=1$, yielding the final retrieved set \(\mathcal{T}_{q}\) with \(|\mathcal{T}_{q}| = 50\).
Baseline retrievers are required to return the top 50 triplets to ensure a comparable retrieval budget across all methods.  
To ensure a fair comparison, {\ours} and all baseline retrievers that depend on external encoders use the same strong text encoder, i.e., Dragon~\citep{lin2023train}. We also use a fixed random seed (42) to eliminate randomness and ensure reproducibility.

All retrievers are evaluated in a zero-shot setting on the question-answering datasets, without any supervised task-specific customization.

\begin{table*}[t]
\centering
\caption{Evaluation results on LC-QuAD.}
\label{tab:lcquad2-nockpt-full}
\resizebox{\textwidth}{!}{%
\begin{tabular}{l|l|cc|c|c|c}
\toprule
\multirow{2}{*}{\makecell[c]{\textbf{Retriever}\\\textbf{Category}}} & \multirow{2}{*}{\textbf{Method}} &
\multicolumn{2}{c|}{\textbf{Retrieval}} &
\multicolumn{3}{c}{\textbf{Generation}} \\
\cline{3-7}
& & \makecell[c]{\textbf{Macro Triplet}\\\textbf{Recall@50}} & \makecell[c]{\textbf{Path}\\\textbf{Recall@50}} &
\textbf{Qwen3-8B} & \textbf{LLaMA2-7B} & \textbf{ChatGPT-3.5} \\
\midrule

\multirow{1}{*}{\makecell[c]{\textbf{\ding{55}}}}
& LLM-only
& -- & --
& 19.95 & 21.40 & 23.10 \\
\midrule

\multirow{2}{*}{\makecell[c]{\textbf{Lexical}\\\textbf{Based}}}
& BM25
& 16.51 & 26.93
& 15.47 & 25.07 & 29.52 \\
& TF-IDF
& 28.12 & 43.36
& 24.84 & 32.83 & 37.09 \\
\midrule

\multirow{9}{*}{\makecell[c]{\textbf{Semantic}\\\textbf{Based}}}
& BCE
& 74.97 & 98.52
& 50.88 & 46.40 & 52.85 \\
& BGE
& 82.46 & 99.31
& 54.23 & 50.72 & 55.44 \\
& ColBERT
& 75.95 & 97.64
& 49.84 & 49.93 & 53.51 \\
& DistilBERT
& 26.27 & 45.33
& 22.64 & 30.87 & 33.65 \\
& Dragon
& 83.14 & \underline{99.34}
& 54.23 & 50.00 & \underline{57.18} \\
& Hybrid
& 83.92 & \underline{99.34}
& \underline{54.65} & 51.87 & \textbf{58.36} \\
& Jina
& 80.59 & 99.21
& 52.52 & 49.54 & 55.28 \\
& MXBAI
& 1.34 & 2.43
& 2.03 & 18.28 & 18.32 \\
& SPLADE
& 80.11 & 98.95
& 52.62 & 49.28 & 53.87 \\
\midrule

\multirow{2}{*}{\makecell[c]{\textbf{Structure}\\\textbf{Based}}}
& G-RAG
& 71.17 & 97.31
& 51.18 & 53.60 & 56.98 \\
& G-Retriever
& 34.98 & 51.43
& 32.34 & 38.83 & 41.09 \\
\midrule

\multirow{3}{*}{\makecell[c]{\textbf{Graph}\\\textbf{Pretraining}}}
& SKP
& 34.28 & 56.41
& 30.14 & 38.56 & 40.92 \\
& \ours ($N=1$)
& \underline{86.91} & \textbf{99.41}
& \textbf{54.82} & \underline{50.75} & \underline{57.18} \\
& \ours ($N=2$)
& \textbf{87.17} & \underline{99.34}
& 53.70 & \textbf{50.85} & 56.82 \\
\bottomrule
\end{tabular}
}
\vspace{-8pt}
\end{table*}





\section{Additional Experiments}
\label{app:exp}

\noindent\textbf{Results on LC-QuAD 2.}
Table~\ref{tab:lcquad2-nockpt-full} evaluates our multi-hop graph retrieval in a graph-RAG pipeline on the additional benchmark LC-QuAD 2. Consistent with results reported in the main experiments, {\ours} improves multi-hop evidence acquisition (triplet- and path-level recall) and yields strong end-to-end answer accuracy with different generators.

\noindent\textbf{Retrieval evaluation.}
{\ours} consistently ranks among the strongest methods on Triplet Recall@50 and Path Recall@50, indicating better coverage of the multi-hop subgraph required for reasoning.The $N=2$ setting achieves the best Macro Triplet Recall@50, while the $N=1$ setting attains the best Path Recall@50, suggesting a trade-off between broader neighborhood exploration and preserving high-quality anchors for full gold-path coverage. 

\noindent\textbf{Generation evaluation.}
In the graph-RAG setting, stronger multi-hop retrieval generally translates into higher generation accuracy, as improved path coverage reduces missing-evidence errors and stabilizes reasoning over the retrieved subgraph. Consistent with this, {\ours} achieves the best (or second-best) accuracy on LC-QuAD 2 for Qwen3-8B and LLaMA2-7B, demonstrating that the proposed anchor-and-expansion retrieval is an effective and portable component for multi-hop graph-based retrieval-augmented generation.

\noindent\textbf{Experiments on One-hop Question Answering}
Although {\ours} is designed for multi-hop retrieval, we additionally evaluate whether it transfers to single-hop question answering. Table~\ref{tab:single} reports results on WebQSP~\citep{yih2016value}, a question-answering dataset where questions typically require only a single supporting triplet. The results indicate that, although {\ours} is tailored for multi-hop retrieval, it still achieves comparable performance on single-hop question answering by reliably retrieving the necessary evidence, enabled by its anchor-triplet retrieval mechanism.



\begin{table}[t]
\centering
\caption{Evaluation results on the WebQSP dataset.}
\label{tab:single}
\resizebox{0.65\columnwidth}{!}{%
\begin{tabular}{l|l|cccc}
\toprule
\multirow{2}{*}{\makecell[c]{\textbf{Retriever}\\\textbf{Category}}} & \multirow{2}{*}{\textbf{Methods}} &
\multicolumn{4}{c}{\textbf{ChatGPT}} \\
\cline{3-6}
& & \textbf{Acc} & \textbf{Prec} & \textbf{Rec} & \textbf{F1} \\
\midrule

\makecell[c]{\textbf{\ding{55}}}
& {LLM-only}
& 46.62 & 65.97 & 39.48 & 49.40 \\
\midrule

\multirow{2}{*}{\makecell[c]{\textbf{Lexical}\\\textbf{Based}}}
& BM25
& 41.84  & 56.39 & 33.47 & 37.23 \\
& TF-IDF
& 47.42 & 60.75 & 35.88 & 40.04 \\
\midrule

\multirow{9}{*}{\makecell[c]{\textbf{Semantic}\\\textbf{Based}}}
& BCE
& 58.49 & 68.92 & 42.18 & 46.76 \\
& BGE
& 59.87 & 68.49 & 42.46 & 46.91 \\
& ColBERT
& 66.95 & 77.70 & 48.01 & 53.07 \\
& DistilBERT
& 39.14 & 57.30 & 34.30 & 42.67 \\
& Dragon
& 63.71 & 73.25 & 44.97 & 49.82 \\
& Hybrid
& 50.38 & 64.25 & 37.52 & 47.37 \\
& Jina
& 58.75 & 68.43 & 42.03 & 46.63 \\
& MXBAI
& 40.25 & 57.62 & 34.69 & 38.45 \\
& SPLADE
& 59.12 & 68.18 & 41.46 & 45.96 \\
\midrule

\multirow{2}{*}{\makecell[c]{\textbf{Structure}\\\textbf{Based}}}
& G-RAG
& 50.53 & 67.69 & 40.64 & 45.24 \\
& G-Retriever
& 52.82 & 68.92 & 42.68 & 47.17 \\
\midrule

\multirow{2}{*}{\makecell[c]{\textbf{Graph}\\\textbf{Pretraining}}}
& SKP
& 43.86 & 60.44 & 35.80 & 39.89 \\
& \ours
& 55.59 & 68.49 & 41.54 & 46.11 \\
\bottomrule
\end{tabular}
}
\end{table}